\documentclass[leqno,twoside,a4paper]{article}
\usepackage{mypreprint,amsmath,amssymb}
\usepackage[british]{babel}
\usepackage[sans]{dsfont}
\newdefin{dfn}{Definition}
\newtheorem{theorem}{Theorem}
\newtheorem{proposition}[theorem]{Proposition}
\newdefin{remark}{Remark}
\newcommand{\Tr}{\operatorname{Tr}}
\newcommand{\norm}[1]{\left\Vert#1\right\Vert}
\newcommand{\rmd}{\mathrm{d}}
\newcommand{\rmi}{\mathrm{i}}

\newcommand{\rmq}{\mathrm{q}}
\newcommand{\rmc}{\mathrm{c}}
\newcommand{\rmf}{\mathrm{f}}
\newcommand{\calG}{\mathcal{G}}
\newcommand{\calH}{\mathcal{H}}
\newcommand{\calL}{\mathcal{L}}
\newcommand{\calT}{\mathcal{T}}
\newcommand{\calI}{\mathcal{I}}
\newcommand{\calJ}{\mathcal{J}}
\newcommand{\calF}{\mathcal{F}}
\newcommand{\calA}{\mathcal{A}}
\newcommand{\calB}{\mathcal{B}}
\newcommand{\calM}{\mathcal{M}}
\newcommand{\calS}{\mathcal{S}}
\newcommand{\openone}{\mathds1}

\begin{document}

\mathclass{Primary 81P15; Secondary 94A17.}

\abbrevauthors{A. Barchielli and G. Lupieri}
 \abbrevtitle{Instruments and entropies}

\title{Instruments and mutual entropies \\
in quantum information}

\author{Alberto Barchielli}
\address{Politecnico di Milano, Dipartimento di Matematica, \\
Piazza Leonardo da Vinci 32, I-20133 Milano, Italy. \\
E-mail: Alberto.Barchielli@polimi.it}
\thanks{Work partially supported by the \emph{European Community's Human Potential
Programme} under contract HPRN-CT-2002-00279, QP-Applications.}

\author{Giancarlo Lupieri}
\address{Universit\`a degli Studi di Milano,
Dipartimento di Fisica,\\ Via Celoria 16, I-20133 Milano, Italy.
\\
E-mail: Giancarlo.Lupieri@mi.infn.it}

\maketitlebcp

\begin{abstract}
General quantum measurements are represented by instruments. In this paper the mathematical
formalization is given of the idea that an instrument is a channel which accepts a quantum
state as input and produces a probability and an a posteriori state as output. Then, by using
mutual entropies on von Neumann algebras and the identification of instruments and channels,
many old and new informational inequalities are obtained in a unified manner. Such
inequalities involve various quantities which characterize the performances of the instrument
under study; in particular, these inequalities include and generalize the famous Holevo's
bound.
\end{abstract}

\section{Introduction.}
The following problem appears in the field of quantum communication and in quantum statistics:
a collection of statistical operators with some a priori probabilities (initial ensemble)
describes the possible initial states of a quantum system and an observer wants to decide  in
which of these states the system is by means of a quantum measurement on the system itself.
The quantity of information given by the measurement is the classical mutual information
$I_\rmc$ of the input/output joint distribution (Shannon information). Interesting upper and
lower bounds for $I_\rmc$, due to the quantum nature of the measurement, are given in the
literature \cite{Hol73,YueO93,Scu95,SchWW96,Hal97,Jac03}, where the measurement is described
by a \emph{generalized observable} or \emph{positive operator valued} (POV) \emph{measure}; an
exception is the paper \cite{SchWW96}, which considers also the information left in the
post-measurement states.

With respect to a POV measure, a more detailed level of description of the quantum measurement
is given by an \emph{instrument} \cite{Dav76,Oza84}: given a quantum state (the preparation)
as input, the instrument gives as output not only the probabilities of the outcomes but also
the state after the measurement, conditioned on the observed outcome (the a posteriori state).
We can think the instrument to be a channel: from a quantum state (the pre-measurement state)
to a quantum/classical state (a posteriori state plus probabilities). The mathematical
formalization of the idea that an instrument \emph{is} a channel is given in Section
\ref{instrsec}, together with a new construction of the a posteriori states. In Section
\ref{ment+bounds}, by using the identification of the instrument with a channel and the notion
of quantum mutual entropy, we are able to give a unified approach to various bounds for
$I_\rmc$ and for related quantities, which can be thought to quantify the informational
performances of the instrument. One of the most interesting inequality is the strengthening
(\ref{SWW}) of Holevo's bound (\ref{Holevo's_bound}); in the finite case it has been obtained
in Ref.\ \cite{SchWW96} where the authors introduce a specific model of the measuring process
(without speaking explicitly of intruments) and use the strong subadditivity of the von
Neumann entropy. The introduction of the general notion of instrument, the association to it
of a channel and the use of Uhlmann's monotonicity theorem allows us to obtain the same result
in a more direct way and to extend it to a more general set up. In Section \ref{Hall} a new
upper bound (\ref{newbound}) for the classical mutual information $I_\rmc$ is obtained by
combining an idea by Hall \cite{Hal97} and inequality (\ref{SWW}).

We already gave some results in \cite{BarL04pr}, mainly in the discrete case. Here we give the
general results, which are based on the theory of relative entropy on von Neumann algebras
\cite{OhyP93}. Continuous parameters appear naturally in quantum statistical problems, but
also in the quantum communication set up infinite dimensional Hilbert spaces and general
initial ensembles are needed \cite{Yue97,HolS04}. Some of the informational quantities
presented here have been studied in \cite{Bar01,BarL04} in the case of instruments describing
continual measurements.

\subsection{Notations and preliminaries.}
\subsubsection{Bounded operators.}
We denote by $\calL(\calA;\calB)$ the space of bounded linear operators from $\calA$ to
$\calB$, where $\calA,\,\calB$ are Banach spaces; moreover we set
$\calL(\calA):=\calL(\calA;\calA)$.

\subsubsection{Quantum states.}
Let $\calH$ be a separable complex Hilbert space; a normal state on $\calL(\calH)$ is
identified with a statistical operator, $\calT(\calH)$ and $\calS(\calH)\subset \calT(\calH)$
are the trace-class and the space of the statistical operators on $\calH$, respectively,  and
$\langle \rho,a\rangle := \Tr_{\calH} \{\rho a\}$, $\rho \in \calT(\calH)$, $a\in
\calL(\calH)$.

More generally, if $a$ belongs to a $W^*$-algebra and $\rho$ to its dual $\calM^*$ or predual
$\calM_*$, the functional $\rho$ applied to $a$ is denoted by $\langle \rho,a\rangle$.

\subsubsection{A quantum/classical algebra.} \label{qcalgebra}
Let $(\Omega,\calF,Q)$ be a measure space, where $Q$ is a $\sigma$-finite measure. By Theorem
1.22.13 of \cite{Sak71}, the $W^*$-algebra $\calL(\calH)\otimes L^\infty(\Omega,\calF,Q)$
($W^*$-tensor product) is naturally isomorphic to the $W^*$-algebra
$L^\infty\big(\Omega,\calF,Q;\calL(\calH)\big)$ of all the $\calL(\calH)$-valued
$Q$-essentially bounded weakly$^*$ measurable functions on $\Omega$. Moreover (\cite{Sak71},
Proposition 1.22.12), the predual of this $W^*$-algebra is
$L^1\big(\Omega,\calF,Q;\calT(\calH)\big)$, the Banach space of all the $\calT(\calH)$-valued
Bochner $Q$-integrable functions on $\Omega$, and this predual is naturally isomorphic to
$\calT(\calH)\otimes L^1(\Omega,\calF,Q)$ (tensor product with respect to the greatest cross
norm --- \cite{Sak71}, pp.\ 45, 58, 59, 67, 68).

Let us note that a normal state $\Sigma$ on $L^\infty\big(\Omega,\calF,Q;\calL(\calH)\big)$ is
a measurable function $\omega \mapsto \Sigma(\omega)\in \calT(\calH)$, $\Sigma(\omega)\geq 0$,
such that $\Tr_{\calH}\{\Sigma(\omega)\}$ is a probability density with respect to $Q$.

\subsubsection{Quantum channels.} A \emph{channel} $\Lambda$ (\cite{OhyP93} p.\ 137), or
dynamical map, or stochastic map is a completely positive linear map, which transforms states
into states; usually the definition is given for its adjoint $\Lambda^*$. The channels are
usually introduced to describe noisy quantum evolutions, but we shall see that also quantum
measurements can be identified with channels.
\begin{dfn}\label{channdfn}
Let $\calM_1$ and $\calM_2$ be two $W^*$-algebras.  A linear map $\Lambda^*$ from $\calM_2$ to
$\calM_1$ is said to be a \emph{channel} if it is completely positive, unital (i.e.\ identity
preserving) and normal (or, equivalently, weakly$^*$ continuous).
\end{dfn}
\begin{remark}
Due to the equivalence of w$^*$-continuity and existence of a preadjoint $\Lambda$
\cite{Dix57}, Definition \ref{channdfn} is equivalent to: $\Lambda$ is a completely positive
linear map from the predual $\calM_{1*}$ to the predual $\calM_{2*}$, normalized in the sense
that $\langle \Lambda[\rho], \openone_2 \rangle_2= \langle \rho,\openone_1\rangle_1$, $\forall
\rho\in \calM_{1*}$. Let us note also that $\Lambda$ maps normal states on $\calM_1$ into
normal states on $\calM_2$.
\end{remark}
\begin{remark} Note that the composition of
channels gives again a channel. If we have three channels $\Lambda^*_1 : \calM_2 \to \calM_1$,
$\Lambda^*_2 : \calM_3 \to \calM_2$, $\Lambda^*_3 : \calM_3 \to \calM_1$ and such that
$\Lambda_2\circ \Lambda_1=\Lambda_3$, following \cite{OhyP93} we say that $\Lambda_3$ is a
\emph{coarse graining} of $\Lambda_1$ or that $\Lambda_1$ is a \emph{refinement} of
$\Lambda_3$.
\end{remark}

\subsection{Entropy.}
\subsubsection{Relative entropies.} The general definition of
the relative entropy $S(\Sigma|\Pi)$ for two states $\Sigma$ and $\Pi$ is given in
\cite{OhyP93}; here we give only some particular cases of the general definition.

Given a separable Hilbert space $\calH$ and two states $ \sigma,\,\tau\in \calS(\calH)$ the
quantum relative entropy of $\sigma $ with respect to $\tau$ is defined by
\begin{equation}\label{relqentropy}
S_\rmq(\sigma|\tau):=\Tr_\calH\{\sigma(\log \sigma-\log \tau)\}.
\end{equation}

Given two normal states $P_i$ on $L^\infty(\Omega,\calF,Q)$, i.e.\ two probability measures
such that $P_i(\rmd \omega)=q_i( \omega) Q(\rmd\omega)$, the classical relative entropy of
$P_1$ with respect to $P_2$, or Kullback-Leibler divergence, is
\begin{equation}
S_\rmc(P_1|P_2):= \int_\Omega Q(\rmd \omega)\,q_1(\omega) \log\frac{q_1(\omega)}{q_2( \omega)}
\equiv \int_\Omega P_1(\rmd \omega) \,\log\frac{P_1(\rmd \omega)}{P_2(\rmd \omega)}\,.
\end{equation}

Given two normal states $\Sigma_k$  on $L^\infty\big(\Omega,\calF,Q;\calL(\calH)\big)$, the
relative entropy of $\Sigma_1$ with respect to $\Sigma_2$ is
\begin{equation}\label{cqS1}
S(\Sigma_1|\Sigma_2) = \int_\Omega Q(\rmd \omega) \Tr_\calH\left\{\Sigma_1(\omega)\big(\log
\Sigma_1(\omega)-\log \Sigma_2(\omega)\big)\right\}.
\end{equation}
Let us define the two probabilities $P_k(\rmd \omega):= \Tr_\calH\{\Sigma_k(\omega)\} Q(\rmd
\omega)$ and the two measurable families of density operators $\sigma_k(\omega):=
\Sigma_k(\omega)/\Tr_\calH\{\Sigma_k(\omega)\}$ (these definitions hold where the denominators
do not vanish and are completed arbitrarily where the denominators vanish). Then, eq.\
(\ref{cqS1}) gives immediately
\begin{equation}\label{cqS2}
S(\Sigma_1|\Sigma_2) = S_\rmc(P_1|P_2) + \int_\Omega P_1(\rmd \omega) \,
S_\rmq\big(\sigma_1(\omega)| \sigma_2(\omega)\big).
\end{equation}

Finally, let us denote by $S_\rmq(\eta)$ \emph{the von Neumann entropy}, i.e.
\begin{equation}
S_\rmq(\eta)= -\Tr_\calH\{\eta \log \eta \}, \qquad \eta\in \calS(\calH).
\end{equation}
All the relative entropies and entropies take values in $[0,+\infty]$. Note that we have used
a subscript ``c'' for classical quantities, a subscript ``q'' for purely quantum ones and no
subscript for general quantities, eventually of a mixed character.

\subsubsection{Convexity properties.}

A key result which follows from the convexity properties of the relative entropy is
\emph{Uhlmann's monotonicity theorem} (\cite{OhyP93}, Theor.\ 1.5 p.\ 21), which implies that
channels decrease the relative entropy.
\begin{theorem}\label{Uhltheo}
If $\Sigma$ and $\Pi$ are two normal states on $\calM_1$ and $\Lambda^*$ is a channel from
$\calM_2\to \calM_1$, then $ S(\Sigma| \Pi)\geq S(\Lambda[\Sigma]| \Lambda[\Pi])$.
\end{theorem}
\begin{remark}\label{restr}
Note also that the operation of restricting the states to some subalgebra is a channel; so, if
$\Sigma^{12}$ and $\Pi^{12}$ are two normal states on $\calM_1\otimes \calM_2$ and
$\Sigma^{k}$ and $\Pi^{k}$ are their restrictions to $\calM_k$, then $S(\Sigma^{12}|
\Pi^{12})\geq S(\Sigma^k| \Pi^k)$, $k=1,2$.
\end{remark}

\subsubsection{Mutual entropies.} The classical notion of mutual entropy can be immediately
generalized to states on von Neumann algebras. Let $\Sigma^{12}$ be a normal state on
$\calM_1\otimes \calM_2$ and let us denote by $\Sigma^{1}$ and $\Sigma^{2}$ its
\emph{marginals}, i.e.\ its restrictions to $\calM_1$ and $\calM_2$, respectively. The
\emph{mutual entropy} of $\Sigma^{12} $ is by definition the relative entropy
$S(\Sigma^{12}|\Sigma^1\otimes\Sigma^2)$ of the state with respect to the tensor product of
its marginals. We shall use the following results on mutual entropies.
\begin{remark} Let $\Sigma^{123}$ be a normal state on $\calM_{1}\otimes\calM_2\otimes\calM_3$
and denote all its possible marginals by $\Sigma^{ij}$ ($i<j$ with $i=1,2$ and $j=2,3$),
$\Sigma^{j}$ ($j=1,2,3$). From Corollary 5.20 of Ref.\ \cite{OhyP93} we obtain the chain rules
\begin{equation}\label{chain}
S(\Sigma^{123}|\Sigma^{1}\otimes\Sigma^{2}\otimes\Sigma^{3})= \begin{cases}
S(\Sigma^{123}|\Sigma^{1}\otimes\Sigma^{23})+ S(\Sigma^{23}|\Sigma^{2}\otimes\Sigma^{3})
\\
S(\Sigma^{123}|\Sigma^{13}\otimes\Sigma^{2})+ S(\Sigma^{13}|\Sigma^{1}\otimes\Sigma^{3})
\\
S(\Sigma^{123}|\Sigma^{12}\otimes\Sigma^{3})+ S(\Sigma^{12}|\Sigma^{1}\otimes\Sigma^{2})
\end{cases}
\end{equation}
and from Remark \ref{restr} we obtain
\begin{equation}
S(\Sigma^{123}|\Sigma^{1}\otimes\Sigma^{23})\geq \begin{cases}
S(\Sigma^{12}|\Sigma^{1}\otimes\Sigma^{2})
\\
 S(\Sigma^{13}|\Sigma^{1}\otimes\Sigma^{3})
\end{cases}
\end{equation}
and the similar inequalities given by permutation of the indices.
\end{remark}

\section{Instruments, channels and a posteriori states.} \label{instrsec}

\subsection{Instruments.}
The notion of instrument is central in quantum measurement theory; an instrument gives the
probabilities and the state changes \cite{DavL70,Dav76,Oza84}.

\begin{dfn}\label{instrdfn}
Let $\calH_1,\, \calH_2$ be two separable complex Hilbert spaces and $( \Omega,\calF)$ be a
measurable space. An  \emph{instrument} $\calI$ is a map valued measure such that
\begin{enum}{(iii)}
\renewcommand\theenumi{\roman{enumi}}
\renewcommand\labelenumi{(\theenumi)}
\item $\calI: \calF \to \calL\big(\calT(\calH_1);\calT(\calH_2)\big)$,
\item  $\calI(F)$ is completely positive, $\forall F\in \calF$,
\item (normalization) $\Tr_{\calH_2}\left\{ \calI(\Omega)[\rho]\right\}=\Tr_{\calH_1}\left\{
\rho\right\}$, $\forall \rho\in \calT(\calH_1)$,
\item ($\sigma$-additivity) for every countable family $\{F_i\}$ of pairwise disjoint sets in
$\calF$
\[
\sum_i \left\langle \calI(F_i)[\rho],\, a \right\rangle_2= \Big\langle \calI\Big(\bigcup_i
F_i\Big)[\rho],\, a \Big\rangle_2\,, \qquad\forall \rho\in \calT(\calH_1), \quad\forall a\in
\calL(\calH_2).
\]
\end{enum}
\end{dfn}

Unlike the usual definitions of instrument we have introduced two Hilbert spaces, an initial
one $\calH_1$ and a final one $\calH_2$; we allow the Hilbert space where the quantum system
lives to be changed by the measurement, which is the standard set up when quantum channels are
considered \cite{OhyP93} and which is usefull when we shall construct something similar to the
compound state of Ohya \cite{Ohy83}.

\begin{remark}
The map $F \mapsto E_\calI(F):=\calI(F)^*[\openone_2]$ turns out to be a positive operator
valued (POV) measure on $\calH_1$ (the observable associated with the instrument $\calI$). For
every $\rho\in \calS(\calH_1)$ the map $F \mapsto P_\rho(F)$, with
\begin{equation}\label{probI}
P_\rho(F):= \left\langle \rho ,E_\calI(F)\right\rangle_1 \equiv \left\langle \rho
,\calI(F)^*[\openone_2]\right\rangle_1 \equiv \Tr_{\calH_2} \{ \calI(F)[\rho]\},
\end{equation}
is a probability measure: $P_\rho(F)$ is the probability that the result of the measurement be
in $F$ when the pre-measurement state is $\rho$. Moreover, given the result $F$, the
post-measurement state is $\big(P_\rho(F)\big)^{-1} \calI(F)[\rho]$.
\end{remark}

\begin{remark}\label{abscont}
It is easy to show that all the measures $P_\rho$, $\rho \in \calS(\calH_1)$, are absolutely
continuous with respect to $P_\xi$, where $\xi$ is any faithful normal state on
$\calL(\calH_1)$. So, we can fix also a $\sigma$-finite measure $Q$ on $(\Omega,\calF)$ such
that all the probabilities measures $P_\rho$ are absolutely continuous with respect to $Q$.
Moreover we complete $(\Omega,\calF, Q)$ and extend the instrument to the extended
$\sigma$-algebra in the same way as ordinary measures are extended (\cite{Bil95} Problem 3.10,
p.\ 49): for any set $A$ in the extended $\sigma$-algebra, there exist $B,C\in \calF$ such
that $A\vartriangle B\subset C$ ($\vartriangle$ is the symmetric difference) with $Q(C)=0$ and
we define $\calI(A)=\calI(B)$. For the extended objects we use the same symbols as for the
original ones. It is always possible to take for $Q$ a probability measure, but it is
convenient to leave more freedom; for instance, in the case of a discrete $\Omega$ one takes
for $Q$ the counting measure or in the case of a measurement of position and/or momentum one
takes for $Q$ the Lebesque measure.
\end{remark}

\subsection{The instrument as a channel.}
From now on $\calH_1,\,\calH_2$ are two separable complex Hilbert spaces, $(\Omega,\calF,Q)$
is a complete $\sigma$-finite measure space, $\calI$ is an instrument as in Definition
\ref{instrdfn} and the associated probabilities (\ref{probI}) are such that
\begin{equation}
P_\rho \ll Q\,, \qquad \forall \rho\in \calS(\calH_1).
\end{equation}
Then, we introduce the $W^*$-algebras
\begin{equation}\label{alg123}\begin{gathered}
\calM_1:=
\calL(H_1)\,, \qquad \calM_2:= \calL(H_2)\,, \qquad \calM_3:= L^\infty
(\Omega,\calF,Q)\,, \\
\calM_{23}:=\calM_2\otimes \calM_3\equiv L^\infty \big(\Omega,\calF,Q;\calL(\calH_2)\big)\,.
\end{gathered}\end{equation}
\begin{theorem} \label{Theochann} Let us set
\begin{equation}\label{Itochann}
\langle \rho ,\, \Lambda_\calI^* [a\otimes f]\rangle_1 := \int_\Omega f(\omega) \langle
\calI(\rmd \omega)[\rho],a\rangle_2\,, \quad \forall \rho \in \calT(\calH_1), \ \forall a \in
\calM_2\,, \ \forall f\in \calM_3\,;
\end{equation}
by linearity and continuity the map $\Lambda_\calI^*$ can be extended to a channel
\begin{equation}
\Lambda_\calI^* : \calM_{23} \to \calM_1\,.
\end{equation}
Viceversa, the instrument $\calI$ is uniquely determined by the channel.
\end{theorem}
\Proof Let us note that by approximating $f$ with simple functions we get from
(\ref{Itochann}) $\langle \rho ,\, \Lambda_\calI^* [a\otimes f]\rangle_1 \leq
\norm{\rho}_{\calT(\calH_1)} \norm{a}_{\calL(\calH_2)} \norm{f}_{L^\infty}$; then, the direct
statement follows by standard arguments. Viceversa, given a channel $\Lambda_\calI^*$, an
instrument $\calI$ is defined by: $\forall F\in \calF$
\begin{equation}
\langle \calI(F)[\rho],a\rangle_2:=\langle \rho ,\, \Lambda_\calI^* [a\otimes 1_F]\rangle_1\,,
\qquad \forall \rho \in \calT(\calH_1), \quad \forall a \in \calM_2\,.
\end{equation}
The $\sigma$-additivity follows from the weak$^*$ continuity of the channel; all the other
properties are more or less evident.
\endproof

\subsection{A posteriori states.} Now, let us consider the preadjoint of the channel we have
constructed
\begin{equation}
\Lambda_\calI : \calT(\calH_1) \to L^1\big(\Omega,\calF,Q;\calT(\calH_2)\big).
\end{equation}
The quantity $\Lambda_\calI[\rho]$ is an equivalence class of Bochner integrable
$\calT(\calH_2)$-valued functions of $\omega$; let $\omega\mapsto \Lambda_\calI[\rho](\omega)$
be a representative. If $\rho\geq 0$, then $\Lambda_\calI[\rho](\omega)\geq 0$, $Q$-a.s., and
in this case we take the representative to be positive everywhere; we asked the completeness
of $Q$ just to have the freedom of making modifications inside null sets without having to
take care of measurability. Moreover, if $\rho$ is normalized, also $\Lambda_\calI[\rho]$ is
normalized. So, we have $\forall \rho \in \calS(\calH_1)$
\begin{gather}
\Lambda_\calI[\rho](\omega)\geq 0\,, \quad \forall \omega\in \Omega\,, \qquad \int_\Omega
\Tr_{\calH_2} \left\{ \Lambda_\calI[\rho](\omega)\right\} Q(\rmd \omega)=1\,,
\\ \label{RNder}
\frac{P_\rho(\rmd \omega)} {Q(\rmd \omega)} =  \Tr_{\calH_2} \left\{
\Lambda_\calI[\rho](\omega)\right\} \qquad \text{(Radon-Nikodim derivative)},
\\
\int_F\Lambda_\calI[\rho](\omega) Q(\rmd \omega)=\calI(F)[\rho]\,, \quad \forall F\in \calF\,,
\qquad \text{(Bochner integral)}.
\end{gather}

Let us normalize the positive trace-class operators $\Lambda_\calI[\rho](\omega)$ by setting
\begin{equation}\label{apstates}
\pi_\rho(\omega):= \begin{cases} \left(\Tr_{\calH_2} \left\{\Lambda_\calI[\rho](\omega)
\right\} \right)^{-1} \Lambda_\calI[\rho](\omega) & \text{if }
\Tr_{\calH_2}\left\{\Lambda_\calI[\rho](\omega) \right\}>0
\\
\tilde \rho \quad \big(\tilde \rho\in \calS(\calH_2)\text{, fixed} \big) & \text{if }
\Tr_{\calH_2}\left\{\Lambda_\calI[\rho](\omega) \right\}=0
\end{cases}
\end{equation}
By eqs.\ (\ref{RNder})--(\ref{apstates}) we have
\begin{equation}\label{intapstates}
\int_F \pi_\rho(\omega) P_\rho(\rmd \omega)=\calI(F)[\rho]\,, \quad \forall F\in \calF\,,
\qquad \text{(Bochner integral)}.
\end{equation}
This construction gives directly the result by Ozawa on the \emph{existence} of a family
\emph{of a posteriori states} \cite{Oza85a,Oza85b}, with the small generalization of the use
of two Hilbert spaces.

\begin{proposition} \label{propapost}
Let $\calH_1,\, \calH_2,\, \calI$ be as above. For any $\rho\in \calS(\calH_1)$ there exists a
$P_\rho$-a.s. unique family of \emph{a posteriori states} $\{\pi_\rho(\omega),\, \omega \in
\Omega\}$ for $(\rho,\,\calI)$, which means that the function $\pi_\rho: \Omega \to
\calS(\calH_2)$ is measurable and that eq.\ (\ref{intapstates}) holds.
\end{proposition}

Theorem \ref{Theochann} and Proposition \ref{propapost} generalize immediately to the case of
$\calL(\calH_1),\, \calL(\calH_2)$ substituted by von Neumann algebras with separable predual;
the separability is needed in the results quoted in Subsection \ref{qcalgebra} and taken from
\cite{Sak71} and which are at the bases of the whole construction.

\section{Instruments, mutual entropies, informational bounds.}\label{ment+bounds}
\subsection{The letter states and the measurement.}
In quantum statistics, the following problem of identification of states is a natural one.
There is a parametric family of quantum states $\rho_\rmi(\alpha)$ (the subscript ``i'' stays
for ``initial''), where $\alpha$ belongs to some parameter space $A$ and it is distributed
with some a priori probability $P_\rmi$. The experimenter has to make inferences on $\alpha$
by using the result of some measurement on the quantum system. In quantum communication
theory, the problem of the transmission of a message through a quantum channel is similar.  A
message is transmitted by encoding the letters in some quantum states, which are possibly
corrupted by a quantum noisy channel; at the end of the channel the receiver attempts to
decode the message by performing measurements on the quantum system. So, one has an alphabet
$A$ and the letters $\alpha \in A$ are transmitted with some a priori probabilities $P_\rmi$.
Each letter $\alpha$ is encoded in a quantum state and  we denote by $\rho_\rmi(\alpha)$ the
state associated to the letter $\alpha$ as it arrives to the receiver, after the passage
through the transmission channel.

Let us give the formalization of both problems; we use the language of the quantum
communication set up. First of all, we have a $\sigma$-finite measure space $(A,\calA,\nu)$;
$A$ is the alphabet and the a priori probabilities for the letters are given by $P_\rmi(\rmd
\alpha)=q_\rmi(\alpha)\nu(\rmd \alpha)$, where $q_\rmi$ is a suitable probability density with
respect to $\nu$. The \emph{letter states} are $\rho_\rmi(\alpha)\in \calS(\calH_1)$ with
$\alpha \mapsto \rho_\rmi(\alpha)$ measurable and the mixture
\begin{equation}\label{iapriori}
\eta_\rmi= \int_A P_\rmi(\rmd \alpha) \,\rho_\rmi(\alpha)\equiv \int_A \nu(\rmd
\alpha)\,q_\rmi(\alpha)\rho_\rmi(\alpha) \in \calS(\calH_1) \qquad \text{(Bochner integral)}
\end{equation}
can be called the \emph{initial a priori state}. One calls $\{P_\rmi,\rho_\rmi\}$ the
\emph{initial ensemble}. It would be possible to take $P_\rmi$ as $\nu$; then,
$q_\rmi(\alpha)=1$. However, it is convenient to distinguish $P_\rmi$ and $\nu$, mainly for
the cases when one has more initial ensembles. Note that $\alpha \mapsto \rho_rmi(\alpha)$ is
nothing but a random variable in the probability space $(A,\calA,P_\rmi)$ with value in
$\calS(\calH_1)$.

Let the decoding measurement be represented by the instrument $\calI$ of the previous section
with the associated POV measure $E_\calI$. By using the notations of Section \ref{instrsec}
and, in particular, the Radon-Nikodim derivative (\ref{RNder}), we can construct the following
probabilities, conditional probabilities and densities: $\forall F\in \calF$, $\forall B\in
\calA$
\begin{gather}\label{probA}
P_{\rmf|\rmi}(F|\alpha):= P_{\rho_\rmi(\alpha)}(F), \qquad \qquad
q_{\rmf|\rmi}(\omega|\alpha):= \frac{P_{\rmf|\rmi}(\rmd \omega|\alpha)}{Q(\rmd
\omega)}=\Tr_{\calH_2}\{\Lambda_\calI[\rho_\rmi(\alpha)](\omega)\},
\\
P_{\rmf}(F):= \int_A  P_{\rmf|\rmi}(F|\alpha)\, P_\rmi(\rmd \alpha)= P_{\eta_\rmi}(F), \qquad
q_{\rmf}(\omega):= \frac{P_{\rmf}(\rmd \omega)}{Q(\rmd
\omega)}=\Tr_{\calH_2}\{\Lambda_\calI[\eta_\rmi](\omega)\},
\\
P_{\rmi\rmf}(\rmd \alpha\times \rmd \omega):= P_{\rmf|\rmi}(\rmd \omega|\alpha)\,P_\rmi(\rmd
\alpha) , \qquad q_{\rmi\rmf}(\alpha,\omega):= \frac{P_{\rmi\rmf}(\rmd \alpha \times \rmd
\omega)}{\nu(\rmd \alpha)\,Q(\rmd \omega)}=q_{\rmf|\rmi}(\omega|\alpha)q_\rmi(\alpha),
\\ \label{probZ}
P_{\rmi|\rmf}(B|\omega):=\frac{P_{\rmi\rmf}(B\times\rmd  \omega)}{P_\rmf(\rmd \omega)}\,,
\qquad \qquad q_{\rmi|\rmf}(\alpha|\omega):= \frac {P_{\rmi|\rmf}(\rmd
\alpha|\omega)}{\nu(\rmd \alpha)} = \frac{q_{\rmi\rmf}(\alpha,\omega)}{q_{\rmf}(\omega)};
\end{gather}
the subscript ``f'' stays for ``final''.

If we apply the measurement, but we do not do any selection on the system, we obtain the
\emph{post-measurement a priori states}
\begin{equation}\label{post-m}
\eta_{\rmf}^{\alpha}:= \calI(\Omega)[\rho_\rmi(\alpha)],\qquad \eta_\rmf:=
\calI(\Omega)[\eta_\rmi]=\int_A P_\rmi(\rmd\alpha)\,\eta_{\rmf}^{\alpha}.
\end{equation}
By applying the definition (\ref{apstates}) we can introduce two families of a posteriori
states:
\begin{equation}\label{2apost}
\rho_{\rmf}^\alpha(\omega):= \pi_{\rho_\rmi(\alpha)}(\omega), \qquad \rho_\rmf(\omega):=
\pi_{\eta_\rmi}(\omega).
\end{equation}
By using eqs.\ (\ref{intapstates}) for $F=\Omega$, (\ref{iapriori})--(\ref{2apost}), one
obtains
\begin{equation}\label{somemixtures}
\begin{gathered}
\int_\Omega
P_{\rmf|\rmi}(\rmd\omega|\alpha)\,\rho_{\rmf}^{\alpha}(\omega)=\eta_{\rmf}^{\alpha}, \qquad
\int_A P_{\rmi|\rmf}(\rmd\alpha|\omega)\,\rho_{\rmf}^{\alpha}(\omega)=\rho_\rmf(\omega),
\\
\int_\Omega P_{\rmf}(\rmd\omega)\,\rho_\rmf(\omega)=\eta_\rmf, \qquad \int_{A\times \Omega}
P_{\rmi\rmf}(\rmd \alpha\times \rmd\omega)\,\rho_\rmf^\alpha(\omega)=\eta_\rmf;
\end{gathered}
\end{equation}
here and in the following integrals on states are in the Bochner sense. Let us stress that the
states $\rho_\rmi(\alpha)$, $\eta_{\rmf}^{\alpha}$ are uniquely defined $P_\rmi$-almost
surely, $\rho_\rmf(\omega)$ $P_\rmf$-a.s.\ and $\rho_{\rmf}^{\alpha}(\omega)$
$P_{\rmi\rmf}$-a.s.

\subsection{Algebras and states.}
With respect to the algebras given in (\ref{alg123}) we have one more von Neumann algebra,
$L^\infty(A,\calA, \nu)$; then, we set
\begin{equation}
\begin{aligned}
&\calM_0 := L^\infty(A,\calA, \nu), \qquad  &\calM_{ij}:=\calM_i\otimes \calM_j\,, \quad i<j,
\\
&\calM_{ijk}:=\calM_{ij}\otimes \calM_k\,, \quad i<j<k, \qquad
&\calM_{0123}:=\calM_{01}\otimes \calM_{23}\,;
\end{aligned}
\end{equation}
in particular, we have the identification
\begin{equation}
\calM_{01}= \calM_0 \otimes \calM_1= L^\infty\big(A,\calA, \nu; \calL(\calH_1)\big).
\end{equation}
The states are represented by densities with respect to $\int_A \nu(\rmd \alpha) \ldots$,
$\Tr_{\calH_1} \{\ldots\}$,  $\Tr_{\calH_2} \{\ldots\}$,  $\int_{\Omega} Q(\rmd \omega)
\ldots$

\subsubsection{The initial state.}
It is easy to see that the initial ensemble $\{P_\rmi,\rho_\rmi\}$ can be seen as a normal
state on $\calM_{01}$. By using a superscript which indicates the algebras on which a state is
acting, we can write
\begin{equation}
\Sigma_\rmi^{01}:=\{q_\rmi(\alpha)\rho_\rmi(\alpha)\}, \qquad
\Sigma_\rmi^0=\{q_\rmi(\alpha)\}, \quad \Sigma_\rmi^1=\{\eta_\rmi\},
\end{equation}
for the initial state and its marginals.

\subsubsection{The final state.}
We already constructed the channel $\Lambda_\calI^* : \calM_{23} \to \calM_1$; by dilating it
with the identity we obtain the \emph{measurement channel}
\begin{equation}
\Lambda^* : \calM_{023} \to \calM_{01}\,, \qquad \Lambda^* := \openone \otimes
\Lambda^*_\calI\,.
\end{equation}

By applying the measurement channel to the initial state we obtain the final state
\begin{equation}\label{defsigmaf}
\Sigma^{023}_\rmf:= \Lambda[\Sigma^{01}_\rmi]= \{q_\rmi(\alpha)
\Lambda_\calI[\rho_\rmi(\alpha)](\omega)\} = \{q_{\rmi\rmf}(\alpha,\omega)
\rho_{\rmf}^{\alpha}(\omega)\},
\end{equation}
whose marginals are
\begin{equation}
\begin{aligned}
& \Sigma^{02}_\rmf= \{q_\rmi(\alpha)\eta_{\rmf}^{\alpha}\}, &\qquad & \Sigma^{03}_\rmf=
\{q_{\rmi\rmf}(\alpha,\omega) \}, &\qquad & \Sigma^{23}_\rmf= \{
q_\rmf(\omega)\rho_{\rmf}(\omega)\},
\\
& \Sigma^{0}_\rmf= \Sigma^{0}_\rmi=\{q_\rmi(\alpha)\}, &\qquad & \Sigma^{2}_\rmf= \{\eta_\rmf
\}, &\qquad & \Sigma^{3}_\rmf= \{ q_\rmf(\omega)\}.
\end{aligned}
\end{equation}
Let us note that
\begin{equation}\label{tensf}
\Lambda[\Sigma^{0}_\rmi\otimes \Sigma^{1}_\rmi]=\Sigma^{0}_\rmf\otimes \Sigma^{23}_\rmf\,.
\end{equation}

\goodbreak

\subsection{Mutual entropies, Holevo's bound and other inequalities.}\nobreak
\subsubsection{$\chi$-quantities.}
Holevo's bound (\ref{Holevo's_bound}) involves a mean quantum relative entropy, which is often
called \emph{Holevo's chi-quantity}, given by
\begin{equation}
\chi\{P_{\rmi},\rho_\rmi\}:= \int_A P_\rmi(\rmd \alpha) \,
S_\rmq(\rho_\rmi(\alpha)|\eta_\rmi).
\end{equation}
In general, given a probability space $(B,\calB,P)$ and a measurable family $\beta\mapsto
\tau(\beta)$ of statistical operators on some Hilbert space $\calH$, the $\chi$-quantity of
the ensemble $\{P,\tau\}$ is defined by
\begin{equation}
\chi\{P,\tau\}:= \int_B P(\rmd \beta) \, S_\rmq(\tau(\beta)|\sigma), \qquad \sigma:= \int_B
P(\rmd \beta) \, \tau(\beta);
\end{equation}
in this definition the set $B$ could be $\calS(\calH)$ itself, see \cite{HolS04} pp.\ 2--4. By
using the definition (\ref{relqentropy}) of the quantum relative entropy and the definition of
von Neumann entropy, when $S_\rmq(\sigma)<\infty$, one has
\begin{equation}\label{altchi}
\chi\{P,\tau\}= S_\rmq(\sigma)-  \int_B P(\rmd \beta) \, S_\rmq\big(\tau(\beta)\big).
\end{equation}

The expressions of the mutual entropies we shall need will contain the $\chi$-quantities
$\chi\{P_{\rmi},\rho_\rmi\}$, $\chi\{P_{\rmi},\eta_{\rmf}^\bullet\}$,
$\chi\{P_{\rmf},\rho_{\rmf}\}$, $\chi\{P_{\rmi\rmf},\rho_{\rmf}^\bullet\}$ and the mean
$\chi$-quantities
\begin{equation}
\int_\Omega P_\rmf(\rmd \omega)\, \chi\big\{P_{\rmi|\rmf}(\bullet|\omega),\rho_{\rmf}^\bullet
(\omega)\big\}= \int_{A\times \Omega} P_{\rmi\rmf}(\rmd \alpha\times \rmd \omega)\,
S_\rmq\big(\rho_{\rmf}^{\alpha}(\omega)|\rho_\rmf(\omega)\big),
\end{equation}
\begin{equation}
\int_A P_\rmi(\rmd \alpha)\, \chi\big\{P_{\rmf|\rmi}(\bullet|\alpha),\rho_{\rmf}^\alpha
\big\}= \int_{A\times \Omega} P_{\rmi\rmf}(\rmd \alpha\times \rmd \omega)\,
S_\rmq\big(\rho_{\rmf}^{\alpha}(\omega)|\eta_{\rmf}^{\alpha}\big);
\end{equation}
the mixtures appearing in these $\chi$-quantities are given by eqs.\ (\ref{iapriori}),
(\ref{post-m}), (\ref{somemixtures}).

\subsubsection{Mutual entropies.}
By using the definitions above and property (\ref{cqS2}), it is easy to compute all the mutual
entropies involving the initial and the final state. First of all we get that Holevo's
$\chi$-quantity is the initial mutual entropy
\begin{equation}\label{initmentr}
S(\Sigma_\rmi^{01}|\Sigma_\rmi^0\otimes \Sigma_\rmi^1)= \chi\{P_{\rmi},\rho_\rmi\}
\end{equation}
and that the mutual entropy involving only the classical part of the final state is the
Shannon input/output classical mutual entropy, i.e.\ the classical information on the input
extracted by the measurement:
\begin{equation}\label{2mentr}
S(\Sigma_\rmf^{03}|\Sigma_\rmf^0\otimes \Sigma_\rmf^{3})= S_\rmc(P_{\rmi\rmf}|P_\rmi\otimes
P_\rmf)=: I_\rmc\{P_\rmi,\rho_\rmi;E_\calI\}.
\end{equation}
Then, the remaining mutual entropies turn out to be
\begin{equation}\label{3mentr}
S(\Sigma_\rmf^{02}|\Sigma_\rmf^0\otimes \Sigma_\rmf^2)= \chi\{P_{\rmi},\eta_{\rmf}^\bullet\},
\qquad S(\Sigma_\rmf^{23}|\Sigma_\rmf^2\otimes \Sigma_\rmf^3)= \chi\{P_{\rmf},\rho_{\rmf}\},
\end{equation}
\begin{equation}\begin{aligned}
&S(\Sigma_\rmf^{023}|\Sigma_\rmf^{03}\otimes \Sigma_\rmf^{2})=
\chi\{P_{\rmi\rmf},\rho_{\rmf}^\bullet\},
\\
&S(\Sigma_\rmf^{023}|\Sigma_\rmf^0\otimes \Sigma_\rmf^{23})=
I_\rmc\{P_\rmi,\rho_\rmi;E_\calI\} + \int_\Omega P_\rmf(\rmd \omega)\,
\chi\big\{P_{\rmi|\rmf}(\bullet|\omega),\rho_{\rmf}^\bullet (\omega)\big\},
\\
&S(\Sigma_\rmf^{023}|\Sigma_\rmf^{02}\otimes \Sigma_\rmf^{3})=
I_\rmc\{P_\rmi,\rho_\rmi;E_\calI\} + \int_A P_\rmi(\rmd \alpha)\,
\chi\big\{P_{\rmf|\rmi}(\bullet|\alpha),\rho_{\rmf}^\alpha\big\}.
\end{aligned}
\end{equation}

\subsubsection{Identities.}
By the chain rules (\ref{chain}) we get
\begin{multline}
S(\Sigma_\rmf^{023}|\Sigma_\rmf^{0}\otimes \Sigma_\rmf^{2}\otimes \Sigma_\rmf^{3}) =
S(\Sigma_\rmf^{023}|\Sigma_\rmf^0\otimes \Sigma_\rmf^{23})+
S(\Sigma_\rmf^{23}|\Sigma_\rmf^2\otimes \Sigma_\rmf^{3})
\\
{}= S(\Sigma_\rmf^{023}|\Sigma_\rmf^{02}\otimes \Sigma_\rmf^{3})+
S(\Sigma_\rmf^{02}|\Sigma_\rmf^0\otimes \Sigma_\rmf^{2}) =
S(\Sigma_\rmf^{023}|\Sigma_\rmf^{03}\otimes \Sigma_\rmf^{2})+
S(\Sigma_\rmf^{03}|\Sigma_\rmf^0\otimes \Sigma_\rmf^{3}),
\end{multline}
which gives the expression of the ``tripartite'' mutual entropy
\begin{equation}
S(\Sigma_\rmf^{023}|\Sigma_\rmf^{0}\otimes \Sigma_\rmf^{2}\otimes \Sigma_\rmf^{3}) =
I_\rmc\{P_\rmi,\rho_\rmi;E_\calI\} + \chi\{P_{\rmi\rmf},\rho_{\rmf}^\bullet\}
\end{equation}
and the identities
\begin{multline}\label{idts}
\chi\{P_{\rmi\rmf},\rho_{\rmf}^\bullet\}= \chi\{P_{\rmf},\rho_{\rmf}\}+ \int_\Omega
P_\rmf(\rmd \omega)\, \chi\big\{P_{\rmi|\rmf}(\bullet|\omega),\rho_{\rmf}^\bullet
(\omega)\big\}
\\
{}= \chi\{P_{\rmi},\eta_{\rmf}^\bullet\}+ \int_A P_\rmi(\rmd \alpha)\,
\chi\big\{P_{\rmf|\rmi}(\bullet|\alpha),\rho_{\rmf}^\alpha\big\}.
\end{multline}

\subsubsection{The generalized Schumacher-Westmoreland-Wootters inequality. \label{HSWW}}
Uhl\-mann's monotonicity theorem (see Theorem \ref{Uhltheo}) and eqs.\ (\ref{defsigmaf}),
(\ref{tensf}) give us the inequality
\begin{equation}
S(\Sigma_\rmi^{01}|\Sigma_\rmi^0\otimes \Sigma_\rmi^1)\geq
S(\Lambda[\Sigma_\rmi^{01}]|\Lambda[\Sigma_\rmi^0\otimes \Sigma_\rmi^1])=
S(\Sigma_\rmf^{023}|\Sigma_\rmf^0\otimes \Sigma_\rmf^{23});
\end{equation}
by eqs.\ (\ref{initmentr}), (\ref{3mentr}) this inequality becomes
\begin{equation}\label{SWW}
\chi\{P_{\rmi},\rho_\rmi\}\geq I_\rmc\{P_\rmi,\rho_\rmi;E_\calI\} + \int_\Omega P_\rmf(\rmd
\omega)\, \chi\big\{P_{\rmi|\rmf}(\bullet|\omega),\rho_{\rmf}^\bullet (\omega)\big\}.
\end{equation}
In \cite{SchWW96} this inequality was found in the discrete case; in \cite{BarL04pr} it was
derived, again in the discrete case, by using relative entropies as here and the general case
was announced. Roughly, eq.\ (\ref{SWW}) says that the quantum information contained in the
initial ensemble $\{P_{\rmi},\rho_\rmi\}$ is greater than the classical information extracted
in the measurement plus the mean quantum information left in the a posteriori states.
Inequality (\ref{SWW}) can be seen also as giving some kind of information-disturbance
trade-off, a subject to which the paper \cite{D'A03}, which contains a somewhat related
inequality, is devoted.

Holevo's bound \cite{Hol73}, generalized to the continuous case in \cite{YueO93}, is
\begin{equation}\label{Holevo's_bound}
I_\rmc\{P_\rmi,\rho_\rmi;E_\calI\} \leq \chi\{P_{\rmi},\rho_\rmi\},
\end{equation}
or, in terms of mutual entropies,
\begin{equation}
S(\Sigma_\rmf^{03}|\Sigma_\rmf^0\otimes \Sigma_\rmf^{3})\leq
S(\Sigma_\rmi^{01}|\Sigma_\rmi^0\otimes \Sigma_\rmi^1).
\end{equation}
The derivation of Holevo's bound given in \cite{YueO93} is based on a measurement channel
involving only the POV measure, not the whole instrument; the fact that inequality (\ref{SWW})
is stronger than Holevo's bound (\ref{Holevo's_bound}) is  a consequence of the fact that our
channel $\Lambda $ is a refinement of the channel used in \cite{YueO93} (see the discussion
given in \cite{BarL04pr}).

By using one of the identities (\ref{idts}), the inequality (\ref{SWW}) can be rewritten in an
equivalent form, which is slightly more symmetric:
\begin{equation}\label{BL1}
I_\rmc\{P_\rmi,\rho_\rmi;E_\calI\} \leq \chi\{P_{\rmi},\rho_\rmi\} +
\chi\{P_{\rmf},\rho_{\rmf}\} - \chi\{P_{\rmi\rmf},\rho_{\rmf}^\bullet\}.
\end{equation}

\subsubsection{A lower bound.} By restriction of the states (see Remark \ref{restr})
we get the inequality
\begin{equation}
S(\Sigma_\rmf^{023}|\Sigma_\rmf^0\otimes \Sigma_\rmf^{23})\geq
S(\Sigma_\rmf^{02}|\Sigma_\rmf^0\otimes \Sigma_\rmf^{2});
\end{equation}
by eqs.\ (\ref{2mentr}) and (\ref{3mentr}) we get \cite{BarL04pr}
\begin{equation}\label{lb}
I_\rmc\{P_\rmi,\rho_\rmi;E_\calI\} + \int_\Omega P_\rmf(\rmd \omega)\,
\chi\big\{P_{\rmi|\rmf}(\bullet|\omega),\rho_{\rmf}^\bullet (\omega)\big\} \geq
\chi\{P_{\rmi},\eta_{\rmf}^\bullet\},
\end{equation}
which says that the classical information extracted in the measurement plus the mean quantum
information left in the a posteriori states is greater than the quantum information left in
the post-measurement a priori states.

All the other inequalities which can be obtained from the final state are also consequences of
inequality (\ref{lb}) and identities (\ref{idts}).

\subsubsection{The generalized Groenewold-Lindblad inequality.}

Given an instrument $\mathcal{I}$ and a statistical operator $\eta$, an interesting quantity,
which can be called the \emph{quantum information gain}, is
\begin{equation}
I_\rmq\{\eta;\mathcal{I}\}:= S_\rmq(\eta) - \int_\Omega P_\eta(\rmd \omega)\, S_\rmq\big(
\pi_\eta(\omega )\big)\,;
\end{equation}
this is nothing but the quantum entropy of the pre-measurement state minus the mean entropy of
the a posteriori states. It is a measure of the gain in purity (or loss, if negative) in
passing from the pre-measurement state to the post-measurement a posteriori states and it
gives no information on the ability of the measurement in identifying the pre-measurement
state, ability which is contained in $I_\rmc$.

By using the expression of a $\chi$-quantity in terms of entropies and mean entropies, as in
(\ref{altchi}), one can see that, when
\begin{equation}
S_\rmq(\eta_\rmi)< +\infty, \qquad \int_\Omega P_\rmf(\rmd \omega)\,
S_\rmq\big(\rho_\rmf(\omega)\big)< +\infty,
\end{equation}
inequality (\ref{SWW}) is equivalent to
\begin{equation}\label{Iqineq}
I_\rmq\{\eta_\rmi;\mathcal{I}\} \geq I_\rmc\{P_\rmi,\rho_\rmi;E_\calI\} + \int_A P_\rmi(\rmd
\alpha)\, I_\rmq\{\rho_\rmi(\alpha);\mathcal{I}\}.
\end{equation}
Here the state $\eta_\rmi$ is given and $\{P_\rmi,\rho_\rmi\}$ has to be thought as any
demixture of $\eta_\rmi$.

An interesting question is when the quantum information gain is positive. Groenewold has
conjectured \cite{Gro71} and Lindblad \cite{Lin72} has proved that the quantum information
gain is non negative for an instrument of the von Neumann-L\"uders type. The general case has
been settled down by Ozawa, who in \cite{Oza86} has proved the following theorem in the case
$\calH_1=\calH_2$. A shorter proof with respect to Ozawa's one is based on inequality
(\ref{Iqineq}) \cite{BarL04pr}.

\begin{theorem}\label{TheoOza}
Let $\calH_1,\, \calH_2$ be two separable complex Hilbert spaces, $( \Omega,\calF)$ be a
measurable space and $\calI$ a completely positive instrument as in Definition \ref{instrdfn}.
Then,
\begin{description}
\item[(a)] the instrument $\mathcal{I}$ sends any pure input state into almost surely pure a posteriori states
\item if and only if
\item[(b)] $I_\rmq\{\eta;\mathcal{I}\} \geq 0$, for all statistical operators
$\eta$ for which $S_\rmq(\eta)<\infty$.
\end{description}
\end{theorem}
\Proof (b) $\Rightarrow$ (a) is trivial: put a pure state $\eta_\rmi$ into the definition and
you get $\displaystyle 0\leq I_\rmq\{\eta_\rmi;\calI\}=  - \int_\Omega P_\rmf( \rmd
\omega)\,S_\rmq(\rho_\rmf(\omega)) \ \ \Rightarrow \ \ S_\rmq(\rho_\rmf(\omega))=0 \quad
P_\rmf\text{-a.s.}\ \ \Rightarrow \ \ \rho_\rmf(\omega)$ is pure $P_\rmf$-a.s.

To see (a) $\Rightarrow$ (b), we take a demixture of $\eta_\rmi$ into pure states; then, by
(a) also the states $\rho_{\rmf}^{\alpha}(\omega)$ are pure and
$I_\rmq\big\{\rho_\rmi(\alpha);\calI\big\}=0$; then, eq.\ (\ref{Iqineq}) gives
$I_\rmq\{\eta_\rmi;\calI\} \geq S(P_{\rmi\rmf}|P_\rmi\otimes P_\rmf) \geq 0$.
\endproof

\subsection{Compound states and lower bounds on $I_\rmc$.} In \cite{Ohy83} Ohya introduced a
notion of compound states which involves the input and output states of a quantum channel.
Taking inspiration from this idea, we are able to produce some inequalities which strengthen a
lower bound on $I_\rmc\{P_\rmi,\rho_\rmi;E_\calI\}$ given by Scutaru in \cite{Scu95}.

First of all we need some new families of statistical operators and the relationships among
them:
\begin{gather}
\epsilon_{\rmi\rmf}(\omega):= \int_A P_{\rmi|\rmf}(\rmd
\alpha|\omega)\,\rho_\rmi(\alpha)\otimes \eta_{\rmf}^{\alpha},
\\ \label{Scustate}
\epsilon_\rmi(\omega):= \Tr_{\calH_2}\{\epsilon_{\rmi\rmf}(\omega)\}= \int_A
P_{\rmi|\rmf}(\rmd\alpha|\omega)\,\rho_\rmi(\alpha),
\\
\epsilon_\rmf(\omega):= \Tr_{\calH_1}\{\epsilon_{\rmi\rmf}(\omega)\}= \int_A
P_{\rmi|\rmf}(\rmd\alpha|\omega)\,\eta_{\rmf}^{\alpha},
\\ \label{Ohyastate}
\eta_{\rmi\rmf} := \int_\Omega P_\rmf(\rmd \omega)\, \epsilon_{\rmi\rmf}(\omega)= \int_A
P_\rmi(\rmd \alpha) \, \rho_\rmi(\alpha) \otimes \eta_{\rmf}^{\alpha},
\\
\Tr_{\calH_2}\{\eta_{\rmi\rmf}\} = \int_\Omega P_\rmf(\rmd
\omega)\,\epsilon_\rmi(\omega)=\eta_\rmi, \qquad \Tr_{\calH_1}\{\eta_{\rmi\rmf}\} =\int_\Omega
P_\rmf(\rmd \omega)\, \epsilon_\rmf(\omega)=\eta_\rmf,
\\
\tau_\rmf(\alpha):= \int_\Omega P_{\rmf|\rmi}(\rmd \omega|\alpha)\, \rho_\rmf(\omega), \qquad
\int_A P_\rmi(\rmd \alpha)\, \tau_\rmf(\alpha)= \eta_\rmf,
\\
\gamma_{\rmi\rmf}:= \int_\Omega P_\rmf(\rmd \omega) \, \epsilon_\rmi (\omega) \otimes
\rho_\rmf(\omega), \qquad \Tr_{\calH_2}\{\gamma_{\rmi\rmf}\} =\eta_\rmi, \qquad
\Tr_{\calH_1}\{\gamma_{\rmi\rmf}\} =\eta_\rmf.
\end{gather}
The state (\ref{Scustate}) has been introduced by Scutaru \cite{Scu95} and the state
(\ref{Ohyastate}) is similar to the compound state introduced by Ohya \cite{Ohy83} for quantum
channels.

Now, let us construct a first compound state on $\calM_{0123}$ and let us give some of its
marginals:
\begin{equation}
\begin{aligned}
&\Pi^{0123}:= \{q_{\rmi\rmf}(\alpha,\omega)\, \rho_\rmi(\alpha)\otimes
\eta_{\rmf}^{\alpha}\},&\quad & \null &\quad &\null \\
&\Pi^{012}=\{q_\rmi(\alpha)\,\rho_\rmi(\alpha)\otimes \eta_{\rmf}^{\alpha}\}, &\quad &\null
&\quad &\Pi^{123}= \{q_\rmf(\omega)\epsilon_{\rmi\rmf}(\omega)\},
\\
&\Pi^{13}= \{q_\rmf(\omega) \epsilon_\rmi(\omega)\}, &\quad &\Pi^{12}= \{\eta_{\rmi\rmf}\},
&\quad &\Pi^{23}= \{q_{\rmf}(\omega)\epsilon_\rmf(\omega)\},
\\
&\Pi^{1}= \{ \eta_{\rmi}\}, &\quad &\Pi^{2}= \{\eta_\rmf\}, &\quad &\Pi^3=\{\eta_\rmi\}.
\end{aligned}
\end{equation}
For this state we have $S(\Pi^{0123}|\Pi^{012}\otimes \Pi^{3})= I_\rmc
\{P_\rmi,\rho_\rmi;E_\calI\}$ and Remark \ref{restr} gives the inequalities
\begin{equation}
S(\Pi^{0123}|\Pi^{012}\otimes \Pi^{3}) \geq S(\Pi^{123}|\Pi^{12}\otimes \Pi^{3}) \geq
\begin{cases} S(\Pi^{13}|\Pi^{1}\otimes \Pi^{3}) \\ S(\Pi^{23}|\Pi^{2}\otimes \Pi^{3})
\end{cases}
\end{equation}
which give
\begin{equation}
I_\rmc\{P_\rmi,\rho_\rmi;E_\calI\} \geq \chi\{P_\rmf,\epsilon_{\rmi\rmf}\} \geq
\begin{cases} \chi\{P_\rmf,\epsilon_{\rmi}\} \\ \chi\{P_\rmf,\epsilon_{\rmf}\}
\end{cases}
\end{equation}
$I_\rmc\{P_\rmi,\rho_\rmi;E_\calI\} \geq \chi\{P_\rmf,\epsilon_{\rmi}\}$ is Scutaru's bound.

Let us give also a second compound state and some of its marginals:
\begin{equation}
\begin{gathered}
\Gamma^{0123}:= \{q_{\rmi\rmf}(\alpha,\omega)\, \rho_\rmi(\alpha)\otimes
\rho_{\rmf}(\omega)\},  \qquad \Gamma^{012}= \{q_\rmi(\alpha)\rho_{\rmi}(\alpha)\otimes
\tau_\rmf(\alpha)\},
\\
\Gamma^{23}= \{q_\rmf(\omega)\rho_\rmf(\omega)\}, \qquad \Gamma^{01}=
\{q_\rmi(\alpha)\rho_\rmi(\alpha)\},   \qquad \Gamma^{2}= \{\eta_\rmf\},
\\
\Gamma^{123}= \{q_{\rmf}(\omega)\, \eta_\rmi(\omega)\otimes \rho_{\rmf}(\omega)\}, \qquad
\Gamma^{1}= \{\eta_\rmi\}, \qquad \Gamma^{12}= \{\gamma_{\rmi\rmf}\}.
\end{gathered}
\end{equation}
As before we get the inequalities
\begin{equation}
S(\Gamma^{0123}|\Gamma^{01}\otimes \Gamma^{23}) \geq \left.
\begin{cases} S(\Gamma^{123}|\Gamma^{1}\otimes \Gamma^{23}) \\ S(\Gamma^{012}|\Gamma^{01}\otimes \Gamma^{2})
\end{cases} \!\!\!\right\} \geq S(\Gamma^{12}|\Gamma^{1}\otimes \Gamma^{2}),
\end{equation}
\begin{equation}
I_\rmc\{P_\rmi,\rho_\rmi;E_\calI\}  \geq \left.
\begin{cases} \chi\{P_\rmf,\epsilon_{\rmi}\} \\ \chi\{P_\rmi,\tau_{\rmf}\}
\end{cases} \!\!\! \right\} \geq S_\rmq(\gamma_{\rmi\rmf}|\eta_\rmi\otimes \eta_\rmf).
\end{equation}

It is possible to obtain these inequalities also by constructing suitable channels and by
using the idea of the refinement of a channel \cite{BarL04pr}.

\section{Hall's upper bound for $I_\rmc$ and generalizations. \label{Hall}} In \cite{Hal97}
Hall exhibits a transformation on the initial ensemble and on the POV measure which leaves
invariant $I_\rmc$ but not the initial $\chi$-quantity and in this way produces a new upper
bound on the classical information. Inspired by Hall's transformation, a new instrument can be
constructed in such a way that the analogous of inequality (\ref{SWW}) produces an upper bound
on $I_\rmc$ stronger than both Hall's and Holevo's ones.

For simplicity in the following we assume that $\eta_\rmi$ has finite von Neumann entropy and
is invertible:
\begin{equation}
\eta_\rmi\in \calS(\calH_1), \qquad S_\rmq(\eta_\rmi)<+\infty, \qquad \eta_\rmi^{-1}\in
\calL(\calH_1).
\end{equation}
All the traces will be over $\calH_1$.

\subsection{A new instrument $\calJ$.} Let us set
\begin{equation}\label{Malpha}
M(\alpha):= \sqrt{q_\rmi(\alpha)}\ \rho_\rmi(\alpha)^{1/2}\eta_\rmi^{-1/2}\,, \qquad
\calG(\alpha)[\tau] := M(\alpha) \tau M(\alpha)^*\,, \quad \forall \tau \in \calT(\calH_1);
\end{equation}
by eq.\ (\ref{iapriori}) the operators $M(\alpha)$ satisfy the normalization condition
\begin{equation}
\int_A \nu(\rmd\alpha)\, M(\alpha)^*M(\alpha)=\openone\,.
\end{equation}
Then, the position
\begin{equation}
\calJ(\rmd \alpha) := \nu(\rmd\alpha)\, \calG(\alpha)
\end{equation}
defines an instrument from $\calT(\calH_1)$ into $\calT(\calH_1)$ with value space
$(A,\calA)$. The instrument $\calJ$ has been constructed by using only the old initial
ensemble $\{P_\rmi, \rho_\rmi\}$. The associated POV measure is
\begin{equation}\label{newE}
E_\calJ(\rmd\alpha) =\nu(\rmd\alpha)\, M(\alpha)^*M(\alpha)= P_\rmi(\rmd\alpha)\,
\eta_\rmi^{-1/2} \rho_\rmi(\alpha)\eta_\rmi^{-1/2}\,.
\end{equation}

Now, we can construct the associated channel and a posteriori states, as in Section
\ref{instrsec}. By looking at eq.\ (\ref{Itochann}) one has immediately
\begin{equation}
\Lambda_\calJ[\tau](\alpha)=\calG(\alpha)[\tau]=M(\alpha) \tau M(\alpha)^*, \qquad \forall
\tau\in \calT(\calH_1)
\end{equation}
and by looking at eq.\ (\ref{apstates}) one has that, for $\rho\in\calS(\calH_1)$,
\begin{equation}\label{apostalpha}
\tilde\pi_\rho(\alpha):= \begin{cases} \left(\Tr \left\{M(\alpha)^*M(\alpha)\rho \right\}
\right)^{-1} M(\alpha)\rho M(\alpha)^* & \text{if } \Tr \left\{M(\alpha)^*M(\alpha)\rho
\right\} >0
\\
\tilde \rho \quad \big(\tilde \rho\in \calS(\calH_1) \big) & \text{if } \Tr
\left\{M(\alpha)^*M(\alpha)\rho \right\} =0
\end{cases}
\end{equation}
is a family of a posteriori states for $(\rho,\calJ)$. Let us stress that $\calJ$ sends pure
states into a.s.\ pure a posteriori states; therefore, by Theorem \ref{TheoOza} one has
\begin{equation}\label{newIq}
I_\rmq\{\rho;\calJ\} \equiv S_\rmq(\rho) - \int_A \Tr\{E_\calJ(\rmd\alpha)\rho\}\, S_\rmq
\big(\tilde \pi_\rho(\alpha)\big)\geq 0, \qquad \forall \rho\in \calS(\calH_1).
\end{equation}

\subsection{A new initial ensemble.} Let $\{\psi_k\}$ be a c.o.n.s.\ of eigenvectors of
$\eta_\rmi$, so that we can write $\eta_\rmi= \sum_k e_k |\psi_k\rangle \langle \psi_k|$, with
$e_k>0$ and $\sum_k e_k=1$. As in Remark \ref{abscont} one can show that the complex measures
$\langle \psi_k|E_\calI(\rmd \omega) \psi_r\rangle$ are absolutely continuous with respect to
$P_\rmf(\rmd \omega)=\Tr\{\eta_\rmi E_\calI(\rmd \omega)\}= \sum_m e_m \langle
\psi_m|E_\calI(\rmd\omega)\psi_m\rangle$; therefore the Radon-Nikodim derivatives $\langle
\psi_k|E_\calI(\rmd \omega) \psi_r\rangle\big/P_\rmf(\rmd \omega)$ exist and the position
\begin{equation}\label{newinistatrig}
\sigma_\rmi(\omega):= \sum_{kr} \sqrt{e_ke_r}\, |\psi_k\rangle \, \frac{\langle
\psi_k|E_\calI(\rmd \omega) \psi_r\rangle}{P_\rmf(\rmd \omega)}\, \langle \psi_r|
\end{equation}
defines a family of statistical operators; in an abbreviated way we write
\begin{equation}\label{newinistat}
\sigma_\rmi(\omega)= \eta_\rmi^{1/2} \, \frac{E_\calI(\rmd \omega) }{P_\rmf(\rmd \omega)}\,
\eta_\rmi^{1/2}\,.
\end{equation}
Now we consider $\{P_\rmf,\sigma_\rmi\}$ as initial ensemble for $\calJ$; note that one gets
\begin{equation}
\int_\Omega P_\rmf(\rmd\omega) \,\sigma_\rmi(\omega)=\eta_\rmi\,.
\end{equation}

Let us consider now Holevo's bound for the new set up:
\begin{equation}
I_\rmc\{P_\rmf,\sigma_\rmi;E_\calJ\} \leq \chi\{P_\rmf,\sigma_\rmi\}.
\end{equation}
The POV measure $E_\calJ$ and the states $\sigma_\rmi(\omega)$ have been constructed just in
order to have
\begin{equation}
\Tr\{E_\calJ(\rmd \alpha)\sigma_\rmi(\omega)\}= P_{\rmi|\rmf}(\rmd \alpha|\omega),
\end{equation}
as it is easy to verify; this implies immediately
\begin{equation}
I_\rmc\{P_\rmf,\sigma_\rmi;E_\calJ\} =I_\rmc\{P_\rmi,\rho_\rmi;E_\calI\}.
\end{equation}
Therefore, we have
\begin{equation}\label{Halldualbound}
I_\rmc\{P_\rmi,\rho_\rmi;E_\calI\} \leq \chi\{P_\rmf,\sigma_\rmi\}\equiv \int_\Omega
P_\rmf(\rmd\omega)\, S_\rmq\big(\sigma_\rmi(\omega)\big| \eta_\rmi\big),
\end{equation}
which is the ``continuous'' version of Hall's bound $\big($eq.\ (19) of \cite{Hal97}$\big)$.
This bound, in the discrete case, is discussed also in Refs.\ \cite{Hal97b,KinR01,Rus02}.

\subsection{The new upper bound for $I_\rmc$.} Having defined a new instrument and not only a
POV measure, we obtain from (\ref{SWW}) the inequality
\begin{equation}
\chi\{P_{\rmf},\sigma_\rmi\}\geq I_\rmc\{P_\rmi,\rho_\rmi;E_\calI\} + \int_A P_\rmi(\rmd
\alpha)\, \chi\big\{P_{\rmf|\rmi}(\bullet|\alpha),\tilde \pi_{\sigma_\rmi(\bullet)}
(\alpha)\big\},
\end{equation}
which gives a stronger bound than Hall's one (\ref{Halldualbound}). In order to render more
explicit this bound, it is convenient to start from the equivalent form (\ref{Iqineq}), which
now reads
\begin{equation}\label{nnn}
I_\rmq\{\eta_\rmi;\calJ\} \geq I_\rmc\{P_\rmi,\rho_\rmi;E_\calI\} + \int_\Omega P_\rmf(\rmd
\omega)\, I_\rmq\{\sigma_\rmi(\omega);\calJ\}.
\end{equation}
By eqs.\ (\ref{Malpha}) and (\ref{apostalpha}) we obtain $\tilde \pi_{\eta_\rmi}(\alpha)=
\rho_\rmi(\alpha)$; together with eqs.\ (\ref{newIq}), (\ref{newE}), (\ref{altchi}), this
gives
\begin{equation}
I_\rmq\{\eta_\rmi;\calJ\} =\chi\{P_\rmi,\rho_\rmi\}.
\end{equation}
Therefore, eq.\ (\ref{nnn}) gives the new bound
\begin{equation}\label{newbound}
I_\rmc\{P_\rmi,\rho_\rmi;E_\calI\} \leq  \chi\{P_\rmi,\rho_\rmi\} - \int_\Omega P_\rmf(\rmd
\omega)\, I_\rmq\{\sigma_\rmi(\omega);\calJ\};
\end{equation}
let us stress that $I_\rmq\{\sigma_\rmi(\omega);\calJ\}\geq 0$ because of eq.\ (\ref{newIq}).
More explicitly, by eqs.\ (\ref{newE}), (\ref{newinistat}), (\ref{newIq}), we have
\begin{equation}
\int_\Omega P_\rmf(\rmd \omega)\, I_\rmq\{\sigma_\rmi(\omega);\calJ\}= \int_\Omega P_\rmf(\rmd
\omega) \, S_\rmq\big(\sigma_\rmi(\omega)\big) - \int_{A\times \Omega} P_{\rmi\rmf} (\rmd
\alpha \times \rmd \omega)\, S_\rmq \big( \tilde \pi_{\sigma_\rmi(\omega)}(\alpha)\big),
\end{equation}
where $\sigma_\rmi(\omega)$ is given by (\ref{newinistat}) and, by eqs.\ (\ref{Malpha}),
(\ref{apostalpha}), (\ref{newinistat}),
\begin{equation}
\tilde \pi_{\sigma_\rmi(\omega)}(\alpha)= \rho_\rmi(\alpha)^{1/2} \, \frac{E_\calI(\rmd
\omega) }{P_{\rmf|\rmi}(\rmd \omega|\alpha)}\,\rho_\rmi(\alpha)^{1/2}\,;
\end{equation}
this last quantity is defined similarly to (\ref{newinistatrig}), by starting from the
diagonalization of $\rho_\rmi(\alpha)$.

Let us stress that the upper bound in (\ref{newbound}) involves the initial ensemble
$\{P_\rmi,\rho_\rmi\}$ and the POV measure $E_\calI$, not the full instrument $\calI$, while
the bound (\ref{SWW}) involves $\{P_\rmi,\rho_\rmi\}$, $E_\calI$ and also the a posteriori
states of $\calI$. Both bounds (\ref{SWW}) and (\ref{newbound}) are stronger than Holevo's
bound (\ref{Holevo's_bound}).

\end{document}